\documentclass{article}

\usepackage[english]{babel}
\usepackage[utf8x]{inputenc}
\usepackage[T1]{fontenc}
\usepackage[numbers,sort]{natbib}
\usepackage{tikz}
\usepackage{setspace}
\usepackage{authblk}
\usepackage{bm}
\usepackage{amssymb}
\usepackage{multirow}
\usepackage{graphicx}
\usepackage[export]{adjustbox}
\usepackage{mathtools}

\usepackage[a4paper,top=3cm,bottom=2cm,left=3cm,right=3cm,marginparwidth=1.75cm]{geometry}

\usepackage{amsmath}
\usepackage{graphicx}
\usepackage[colorinlistoftodos]{todonotes}
\usepackage[colorlinks=true, allcolors=blue]{hyperref}

\author{Oliver Watson$^{1,*}$, Isidro Cortes-Ciriano$^{2}$, James A Watson$^{3,4}$}
\date{}

\title{A semi-supervised learning framework for quantitative structure-activity regression modelling}

\begin{document}
\maketitle

\noindent 
1: Evariste Technologies Ltd, Goring on Thames, RG8 9AL, United Kingdom \\
2: Centre for Molecular Informatics, Department of Chemistry, University of Cambridge, Lensfield Road, Cambridge, CB2 1EW, United Kingdom.\\
3: Mahidol-Oxford Tropical Medicine Research Unit, Faculty of Tropical Medicine, Mahidol University, Thailand \\
4: Centre for Tropical Medicine and Global Health, Nuffield Department of Medicine, University of Oxford, United Kingdom \\
Correspondence: owatson79@evartech.co.uk

\begin{abstract}
Supervised learning models, also known as quantitative structure-activity regression (QSAR) models, are increasingly used in assisting the process of preclinical, small molecule drug discovery. The models are trained on data consisting of a finite dimensional representation of molecular structures and their corresponding target specific activities. These models can then be used to predict the activity of previously unmeasured novel compounds.
In this work we address two problems related to this approach. 
The first is to estimate the extent to which the quality of the model predictions degrades for compounds very different from the compounds in the training data. 
The second is to adjust for the screening dependent selection bias inherent in many training data sets. In the most extreme cases, only compounds which pass an activity-dependent screening are reported.  
By using a semi-supervised learning framework, we show that it is possible to make predictions which take into account the similarity of the testing compounds to those in the training data and adjust for the reporting selection bias.
We illustrate this approach using publicly available structure-activity data on a large set of compounds reported by GlaxoSmithKline (the Tres Cantos AntiMalarial Set) to inhibit \textit{in vitro} \textit{P. falciparum} growth. 
\end{abstract}

\section{Introduction}

High-throughput experiments allow for the characterisation of the target specific activity of thousands to hundreds of thousands of small molecules \cite{Martis2011,Phatak2009}. The structure-activity data generated from these experiments can be used to fit supervised learning models with the aim of then finding molecular structures that are optimised to maximise multiple desired outcomes, such as target activity, cytotoxicity, and lipophilicity \cite{Cherkasov2014}.
However, many structure activity datasets have an inherent bias in that only molecules with a certain minimal target specific activity are characterised and reported, e.g. \cite{Gamo2010}, or few highly potent molecular structures are reported \cite{Li2010}. A bias towards active molecules will result in overly optimistic predictions of the activity values of new molecules, whereas models trained on data sets mostly comprising inactive molecules might hamper the discovery of structurally novel active compounds \cite{Cortes-Ciriano2018,Norinder2017,Sun2017}. In addition, it is well known that the quality of the predictive model degrades as the similarity between the training and testing compounds decreases \cite{netzeva:05,Wallach2018,Sheridan2015}. The set of testing compounds for which the predictive value of the model is high is known as the applicability domain of the model \cite{netzeva:05}. This is often taken into account by completely restricting models to the domain of compounds similar to those in the training set \cite{netzeva:05}. For a general predictive purposes however, we need ideally to produce a sensible answer for any input compound \cite{Cortes-Ciriano2019}.

In this work, we show that it is possible to explicitly adjust model predictions for both the activity dependent selection bias, and for the distance dependent predictive degradation by accounting for the underlying geometry of molecular space. We use the Tanimoto distance as a metric on molecular space, which has proved a suitable metric to quantify molecular similarity in multiple drug discovery applications \cite{Bajusz2015}. The adjusted predictions are made using a semi-supervised learning framework. This takes as input a set of labelled compounds (structures with labelled activity values) and a larger set of unlabelled compounds (only structures) which provide an empirical representation of the overall distribution of `feasible' small molecules, that is amenable to synthesis and displaying drug-like properties \cite{druglikeness}.
Semi-supervised learning refers to the set of methods developed in machine learning that use labelled (in this case structures with corresponding activity values) and unlabelled data (no corresponding activity values) to build predictive models, see for example \cite{kall:07,shi:11}. The unlabelled data allow for a more accurate representation of the set of `feasible' compounds that could have been part of the (unknown) screening process.

We illustrate this methodology on the Tres Cantos AntiMalarial Set (TCAMS), an open access screening dataset based on the \textit{Plasmodium falciparum} 3D7 asexual assay generated by GlaxoSmithKline (GSK) \cite{Gamo2010}. We fit random forest and ridge regression models to these data. We use held-out data to compare the performance of the semi-supervised framework - which uses the unlabelled data and explicitly adjusts for Tanimoto distance between testing and training data - against the standard fully supervised framework.

\section{Methods}
\subsection{Theoretical framework}
\subsubsection{Overview}

It is generally considered that QSAR models provide reliable predictions for compounds similar to those used as training data, but highly unreliable predictions otherwise \cite{netzeva:05,Cortes-Ciriano2019,Sheridan2012}. 
Our work attempts to formalise this intuition mathematically, assess the evidence for it, and build models that can correct for this effect.

This `generalisability' problem is illustrated in two ways by the TCAMS dataset, a large publicly available set of compounds with measured IC$_{50}$ values of the \textit{in vitro} asexual activities against \textit{P. falciparum} 3D7.
These data only contain screened compounds that inhibited 3D7 growth by more than 80\% at $2\mu$M concentration. This introduces a selection bias for QSAR modelling, as only compounds with considerable activity are reported.
To correct for this bias we are assisted by two sources of information.  Firstly, the original TCAMS publication reported the number of compounds which were screened, so we can estimate the fraction of compounds which passed this screening threshold.
Secondly, we can use available data on a very large number of known compounds that approximately span known feasible compounds showing drug-like properties. These compounds provide a rough approximation of the unknown compounds screened by GSK. In this work we use 2 million structures provided by Molport (a compound vendor).
These structures are unlabelled, i.e. they do not have an assigned activity value.
However, we can use these structures to estimate the probability that a compound is active against \textit{P. falciparum} as a function of its distance to the set of known (active) compounds. The following presents intuition behind the methodology.

For a compound that is entirely different to any of those in the training set of actives, its activity value can be predicted using Bayes rule conditioning on the probability that it is active. The probability that an entirely distinct compound is active is the background rate of actives (0.7\% in the TCAMS data). If it is active, then a reasonable prediction for the activity is the mean observed activity of the active compounds. 
For a compound that is identical to one in the training set, the prediction should approximately be the value corresponding to that training compound. 
The question then is how to interpolate for compounds that fall in between these two scenarios.
We therefore wish to characterise how the mean activity changes as a function of the distance from the training data.  

The intuition behind our approach can be understood through the following analogy. Suppose we have a map where the observed activity value for a given point on the map is the altitude above sea level. Suppose we want to estimate how ``jagged'' the terrain is, where jagged measures how quickly altitude changes between neighbouring points. Suppose further that many observations have been made uniformly at random across the map, but only those with an altitude greater than a given threshold were recorded. 
If the map is blank this implies there is no information content as to the jaggedness of the terrain. If the recorded points are clustered together, this implies that the terrain is divided into low and high regions; in other words, altitude varies smoothly. 
If on the other hand the recorded points are not distinguishable from a set of points chosen uniformly at randomly on the map, this would indicate extremely jagged terrain. 
In our context the Tanimoto distance puts all compounds onto a finite dimensional space corresponding to this map. The unlabelled compounds are used to estimate the data generating process, i.e. an estimate of how compounds are sampled across the `map'. This sampling procedure is very different from a uniform distribution. By comparing the pairwise distances between the active compounds (recorded points) to the pairwise distances between `random' compounds (unlabelled data), we can estimate how smoothly the activity varies as a function of distance to the active compounds.

\subsubsection{Definitions and notation}

We use the following notation throughout. Compounds (small molecules) are denoted $x \in \mathcal{X}$, where $\mathcal{X}$ is the unknown space of all feasible compounds. A compound $x$ is represented by its `fingerprint', a binary vector of some fixed dimension $p$. This vector representation of $x$ is constructed via a `fingerprint mapping', which takes a fixed set of $p$ chemical substructures and determines whether or not these are substructures of $x$. 
In the resulting binary vector, a `1' at index $i$ indicates that the $i$\textsuperscript{th} substructure is a substructure of $x$, and a `0' indicates that it is not. 
It is worth noting that fingerprint mappings are not injective: two different compounds can have the same fingerprint, and there are examples of this in the data we analyse \cite{Rogers2010,Cortes-Ciriano2018}.

Identifying compounds by their fingerprint representation (for a given fingerprint mapping) allows us to define a metric over molecular space. We use the Tanimoto distance (also known as the Jaccard distance), defined as one minus the Tanimoto similarity \cite{Bajusz2015}. The Tanimoto similarity of two compounds $x_i$ and $x_j$ is the number of substructures common to both compounds, divided by the total number of substructures that appear in at least one of the compounds \cite{Bajusz2015}. Written as boolean operators on binary vectors this is $|x_i \cap x_j| / |x_i \cup x_j|$. 
The rationale for choosing this metric is that only sharing a particular substructure provides information regarding similarity, and two compounds that share no substructures are thought of as being maximally different (for want of a better model for representing molecules in a finite dimensional space). 
We denote the Tanimoto distance between compounds $x_i, x_j$ as $d(x_i, x_j)$. For notational simplicity, we do not include the dependency on the underlying fingerprint mapping. This mapping will affect the distance $d$, for example increasing the number of structures $p$ would increase the granularity of $d$. 
In addition we define the setwise Tanimoto distance between a compound $x$ and a set of compounds $\mathcal{S}$ as $d(x, \mathcal{S}) = \text{min}_{s \in \mathcal{S}} d(x, s)$. This is the Tanimoto distance between $x$ and its nearest neighbour in $\mathcal{S}$.

Our semi-supervised structure-activity regression modelling framework applies to the following set-up, whereby there are two distinct sources of data. 
First, we have labelled structure-activity data denoted $\mathcal{L}_n = \{\boldmath{x}_i, \boldmath{y}_i\}_{i=1}^n$ (L for labelled), comprising $n$ compounds. $y_i$ is the response value for the compound $x_i$. In our setting, $y_i$ is the target-specific activity of the compound $x_i$ for some pre-defined target of interest, but in general it could represent other outcomes of interest (e.g. \textit{in vitro} cytotoxicity, or lipophilicity). The response $y_i$ is a (unknown) function of $x_i$ and as such can be written $y_i = y(x_i)$.  
In addition, the responses $y_i$ are all greater than a known cutoff value $L_{\min}$.
We denote as `actives' the molecules with a response value above the $L_{\min}$, and as `inactives' those below the $L_{\min}$. The unknown set of all active molecules is denoted $\mathcal{A}$.  The compounds $x_i$ in our structure-activity dataset $\mathcal{L}_n$ are a strict subset of $\mathcal{A}$ as they have been selected on the basis of observed activities $y_i > L_{\min}$. 
The set $\mathcal{L}_n$ was derived by screening a larger set of compounds $\mathcal{L}_{n'}$ (of known or unknown size $n' > n$), and then choosing the active compounds amongst them: $\mathcal{L}_n = \mathcal{L}_{n'} \cap \mathcal{A}$. The critical point here is that the inactive compounds in the larger set $\mathcal{L}_{n'}$ are unknown or unavailable for analysis.

Second, we have unlabelled structure data of size $N$ denoted $\mathcal{U}_N$ (U for unlabelled). By construction, there are no labelled compounds in $\mathcal{U}_N$ ($\mathcal{U}_N \cap \mathcal{L}_n = \emptyset$).
In general, in this set-up it is assumed that $n << N$, which is that of many semi-supervised learning problems whereby there is a smaller, well curated labelled data set, and a much larger unlabelled data set.

The key assumption that guides the following methodology is that the unlabelled data $\mathcal{U}_N$ are sampled from the same data generating process as the unknown set of screened compounds $\mathcal{L}_{n'}$. It is worth noting that if we knew the structures in $\mathcal{L}_{n'}$ then much of the framework developed here would be unnecessary, but in practice the availability of large sets of active and inactive compounds for a target of interest is rather limited \cite{Cortes-Ciriano2018}, thus strongly limiting predictive modelling applications in preclinical drug discovery. 
We also note that this assumption is, in general untestable, however we show how specific deviations can be detected and corrected for.

\subsubsection{Prediction goal of semi-supervised framework}

Using the two data sources $\mathcal{L}_n$ and $\mathcal{U}_N$, we wish to determine the ranking of the individual molecules in $\mathcal{U}_N$ based on their probabilities of having an activity greater than some pre-specified threshold of interest $I$. 
For example, this threshold could represent an activity high enough to warrant further experiments. We note that in general a ranking based on tail probabilities (function of the mean and higher moments of the distribution) will differ from a ranking based on mean predicted values.
Therefore, for each molecule $x^* \in \mathcal{U}_N$, the goal is to estimate the probability that its activity $y^*$ is greater than the pre-specified threshold value $I$ (where $I$ is significantly greater than $L_{\min}$). 
To estimate this probability, our modelling framework uses the labelled data $\mathcal{L}_n$ to fit a predictive model of $y$ given $x$, using the fingerprint representation of $x \in \mathcal{L}_n$ as a $p$-dimensional predictive variable. 
We use Bayes rule, with the additional knowledge of the background frequency of active compounds, to adjust for
the inherent bias consequent to regressing onto an unrepresentative sample of compounds. In addition, we use the unlabelled data to assess how the conditional probability of being an active compound varies over molecular space. These adjustments are
necessary for the following reasons:

\begin{enumerate}
    \item By construction, all the responses $y_i \in \mathcal{L}_n$ have values greater than $L_{\min}$. Therefore, by regression to the mean, a general regression model will predict for any new compound a value greater than $L_{\min}$, regardless of the overall frequency of active compounds under the data generating process (approximated by $n'/n$).
    \item Using our metric $d$, we can observe whether the active compounds $\mathcal{L}_n$ are closer together than compounds drawn from the same data generating process without selection bias.  Assuming that $\mathcal{L}_n$ were generated by taking the active compounds from a much larger set of compounds generated from the same process that generates the unlabelled data, we can use the inter-compound distances of $\mathcal{L}_n$, compared to inter-compound distances of compounds from $\mathcal{U}_N$ to estimate the rate at which the probability of being active varies as function of distance to the training data under the metric $d$.
\end{enumerate}

Point 1 implies that it is necessary to adjust predictions with the background frequency of active molecules; point 2 implies that a metric on molecular space along with the unlabelled data $\mathcal{U}_N$ provide key additional information as to whether a given molecule $x^*$ is active or not. Specifically, we can use the information on the distance between $x^*$ and the training data $\mathcal{L}_n$ to inform the prediction of $y^*$.

The prediction goal is expressed as the estimation of:
\begin{equation}
    P\left[y^* \geq I |  d(x^*, \mathcal{L}_n) \right]
\end{equation}

By the law of total probability, conditioning on whether $x^*$ is active (i.e. $y^* > L_{\min}$):

\begin{equation}\label{AnalysisGoal}
    P\left[y^* \geq I |  d(x^*, \mathcal{L}_n) \right] = P\left[y^* \geq I |  d(x^*, \mathcal{L}_n) , x^*\in\mathcal{A} \right] P\left[x^*\in \mathcal{A} |  d(x^*, \mathcal{L}_n) \right]
\end{equation}

The omitted second half of the sum with $P\left[y^* \geq I |  d(x^*, \mathcal{L}_n) , x^*\notin\mathcal{A} \right]$, is equal to 0 as, by definition, $y^*$ cannot be greater than $I$ if $x^*$ is not in $\mathcal{A}$.

In the next sections, we outline (i) the estimation of the distance dependent probability that $x^*$ is active: $P\left[x^*\in \mathcal{A} | d(x,\mathcal{L}_n)\right]$; and (ii) the estimation of the conditional probability that $y^* >I$: $P\left[y^* \geq I | d(x^*,\mathcal{L}_n), x^*\in\mathcal{A} \right]$.
We simplify the estimation of (ii) by breaking it down into the predicted expected value of $y^*$, and the predicted uncertainty around this expected value. Assuming a given parametric form for the predictive distribution of $y^*$, we can estimate $P\left[y^* \geq I | d(x^*,\mathcal{L}_n), x^*\in\mathcal{A} \right]$. This can be done by fitting a predictive distribution (conditional on being active) using the active data we have - as explained in a section below - and then re-centering and re-scaling using the mean and variance estimates from the predictive distribution.

\subsubsection{Distance dependent probability that $x^*$ is active}

Applying Bayes rule:

\begin{align}
    P\left[x^*\in \mathcal{A} | d(x^*,\mathcal{L}_n)\right] &= \frac{P\left[x^* \in \mathcal{A}, d(x^*, \mathcal{L}_n) \right]}{P\left[d(x^*, \mathcal{L}_n) \right]} \\
    &= \frac{P\left(x^*\in \mathcal{A}\right) P\left[d(x^*, \mathcal{L}_n) | x^* \in \mathcal{A}\right]}{P\left[d(x^*, \mathcal{L}_n)\right]}
    \label{eq:distance_dependent_activityprob}
\end{align}

We estimate equation \ref{eq:distance_dependent_activityprob} by estimating each of its three components.

First, $P\left[d(x^*, \mathcal{L}_n) | x^* \in \mathcal{A}\right]$ can be estimated via a $v$-fold `cross-prediction' type procedure. For example, taking $v=2$, we can randomly partition $\mathcal{L}_n$ into 2 equally sized subsets $\mathcal{L}^{1}_{n/2}, \mathcal{L}^{2}_{n/2}$. This partition gives a total of $n$ setwise distances for each element of $\mathcal{L}^{1}_{n/2}$ to the set $\mathcal{L}^{2}_{n/2}$, and vice versa. 
By repeating this procedure $k$ times, we obtain $kn$ setwise distances which form an empirical distribution of $P\left[d(x^*, \mathcal{L}_{n/2}) | x^* \in \mathcal{A}\right]$. 
The choice of $v$ corresponds to a bias-variance trade-off. Taking $v=n$ (a leave-one-out procedure) results in $n$ datasets that are likely to be highly similar to one another, resulting in an empirical distribution of $P\left[d(x^*, \mathcal{L}_{n-1}) | x^* \in \mathcal{A}\right]$ with high variance. Lower values of $v$ (e.g. $v=2$) de-correlate the sets used to estimate these setwise distances and result in a lower variance but with increased bias due to the smaller sample sizes.

Second, the denominator $P\left[d(x^* ,\mathcal{L}_n)\right]$ can be estimated using the empirical distribution of setwise distances $d(x, \mathcal{L}_{n})$, where $x \in \mathcal{U}_N$. A sensitivity analysis with respect to the size of the set $\mathcal{L}_n$ can be done by random samples of size $n/2$ elements from $\mathcal{L}_{n}$.

Third, the marginal (prior) $P(x^* \in \mathcal{A})$, which is the overall fraction of active compounds in $\mathcal{X}$, can be estimated in two possible ways. If the number of compounds screened in order to generate the data set $\mathcal{L}_n$ is known, then $n$ over the number of compounds screened approximates the overall fraction of actives in $\mathcal{X}$. 
Otherwise, it is possible to use a limit argument. We assume that compounds very close to an active compound are themselves active: formally this means that $\lim_{d(x^*,\mathcal{L}_n) \to 0} P\left[x^*\in \mathcal{A} | d(x^*,\mathcal{L}_n)\right] = 1$. 
Therefore:
\begin{equation}
    P\left(x^*\in \mathcal{A}\right) = \lim_{d(x^*,\mathcal{L}_n) \to 0}\frac{P\left[d(x^*, \mathcal{L}_n)\right]}{P\left[d(x^*, \mathcal{L}_n) | x^* \in \mathcal{A}\right]}
    \label{limit_ratio}
\end{equation}

This relies on the ability to accurately estimate both terms in the ratio in equation \ref{limit_ratio}. We discuss this in section \ref{stats_estimation}.

\subsubsection{Distance-dependent degradation of predictive accuracy}

In this section we show how to estimate the mean and variance of the predicted value of $y^*$ as a function of the distance between $x^*$ and $\mathcal{L}_n$, conditional on $x^*\in\mathcal{A}$. After fitting a model $M$ to the labelled data $\mathcal{L}_n$, instead of using the `naive' predicted expected value $M(y^* | \mathcal{L}_n)$ (and modelled uncertainty around this estimate), we formally account for degradation in predictive accuracy as a function of the distance $d(x^*, \mathcal{L}_n)$. 
By estimating this distance dependent decrease in model accuracy we can correctly penalise model predictions to obtain a calibrated estimate of $P\left[y^* \geq I | d (x^*, \mathcal{L}_n), x^*\in\mathcal{A} \right]$.

For a given distance $\delta\in[0,1]$, we assess the ability of our predictive model $M$ to extrapolate at a distance $\delta$ from the training data by doing the following:
\begin{itemize}
    \item We standardise the response values $y_i$ so that the model $M$ is fit to approximately standard normal data.
    \item For each compound $x_i\in\mathcal{L}_n$, we construct a subset of the labelled data, defined as all compounds at least $\delta$ units of distance from $x_i$. This is denoted $\bar{\mathcal{L}}_{i,\delta} = \left\{x\in \mathcal{L}_n : d(x, x_i) \geq \delta \right\}$. This is the complement of the $\delta$-ball centred around $x_i$.
    \item We fit the model $M$ to the data $\bar{\mathcal{L}}_{i,\delta}$ and compute the out-of-sample prediction $\hat{y}_{M_{i,\delta}} = M(x_i | \bar{\mathcal{L}}_{i,\delta})$.
\end{itemize}

Here, $M(a|B)$ denotes the prediction on compound $a$ of the model $M$ fit to data $B$.
The $\delta$-distance prediction `quality' of the model $M$ can be assessed by the set of residuals $\{y_i - \hat{y}_{M_{i,\delta}}\}_{i=1}^n$.
The decrease in predictive ability as a function of the setwise distance to the training data can be quantified by estimating smooth functionals $\hat{\beta}(\delta), \hat{\epsilon}(\delta)$, whereby:
\begin{equation} 
    y_i \sim N\left( \hat{\beta}(\delta) \hat{y}_{M_{i,\delta}} , \hat{\epsilon}(\delta)^2\right)
    \label{penalty}
\end{equation}
The estimated standard deviation $\hat{\epsilon}(\delta)$ can be interpreted as 1 minus the distance-$d$ R-squared of the model $M$.
The conditional predictive distribution of the response $y^*$ can then be estimated as: 

\begin{equation}\label{conditionalpredictivedist}
    y^* \sim N\left( \hat{\beta}[d(x^*, \mathcal{L}_n)] M(x^* | \mathcal{L}_n), \hat{\epsilon}[d(x^*, \mathcal{L}_n)] \right)
\end{equation}

\subsubsection{Non-parametric estimation of distance dependent activity covariance}

This section provides a non-parametric method for estimating the distance-dependent covariance of the activity of two compounds. 
This can be used in two ways. First as a general approach for the assessment of the `quality' of a given $p$-dimensional fingerprint mapping. Second, as a conservative estimator for the variance component in equation \eqref{conditionalpredictivedist}.

In general, for any two compounds $x_i,x_j$, the joint distribution of their respective activities $y_i,y_j$ can be estimated as $\left(
 \begin{matrix}
  \bar{y} & \sigma  \\
  \sigma & \bar{y} 
 \end{matrix}\right)
$, where $\bar{y}$ is the mean activity value, and $\sigma$ is the covariance.
If the distance metric over the fingerprint mapping of molecular space is a good representation of the true distance between molecules (and therefore the true average difference in activities), then this covariance $\sigma$ will be a function of the distance $d( x_i , x_j)$ and should be modelled accordingly. 

With this aim, we define $B_{\delta} \subset \mathcal{L}_n\times\mathcal{L}_n$ as the set of all distinct pairs of active compounds for which the pairwise distance is exactly $\delta$:
\begin{equation}
    B_{\delta} = \{ x=(x_i,x_j) : \quad d( x_i ,x_j) = \delta, x_i \neq x_j\}
\end{equation}
The set $B_{\delta}$ can then be used to empirically estimate the distance-dependent covariance function $\sigma^2(\delta)$:
\begin{equation}\label{eq:active_sigma}
    y(x_i) - y(x_j) \sim \text{N}\left[0, \sigma(\delta)^2 \right], \quad x=(x_i,x_j) \in  B_d
\end{equation}
where N is the normal distribution. 

\subsection{Statistical methods}
\subsubsection{Data}

To illustrate our predictive framework, we used the Tres Cantos Antimalarial Set (TCAMS) \cite{Gamo2010} as the labelled data $\mathcal{L}_n$.
These data comprise 13,533 compounds, selected on the basis that they inhibited the growth of \textit{Plasmodium falciparum} 3D7 by at least 80\% at 2 $\mu$M concentration (in this context, this is the assay defining `active' compounds and the threshold $L_{\min}$). This set of compounds was discovered by screening a library of 1,985,056 compounds (an active discovery rate equal to 0.68\%) \cite{Gamo2010}. The structures for the inactive compounds were not reported, and hence, the available structures correspond to only active compounds.

We constructed (see section \ref{stats_estimation}) unlabelled datasets $\mathcal{U}_N$ with publicly available data from the Molport database after having removed all compounds with recorded activities in TCAMS (there were 2044 compounds in Molport with canonical fingerprints equal to compounds in TCAMS, which we count as identical in this case). This gave a total of $N=7,228,997$ compounds with no activity values (unlabelled).

\paragraph{Specific data issues in Molport dataset}

The key assumption used in the estimation of equation \ref{eq:distance_dependent_activityprob} is that the set 
$\mathcal{U}_N$ is sampled from the same data generating process as the unknown set $\mathcal{L}_{n'}$. This allows us to use $\mathcal{U}_N$  to adjust for the inherent selection bias when training a supervised regression model on $\mathcal{L}_n$.

The set of unlabelled data $\mathcal{U}_N$ was provided with a certain ordering (a set of numbered files, each with approximately 500,000 compounds).
This ordering was strongly correlated with the setwise distance to the 13,533 compounds in the TCAMS dataset (labelled data).
The MolPort company could not provide a reason for this particular ordering of their data. 
It would seem likely that the database was compiled over time, and thus the earlier compounds in the list are those that are simpler to synthesise and thus more likely to appear in other high compound collections. 

\begin{figure}
\centering
\includegraphics[width=\linewidth]{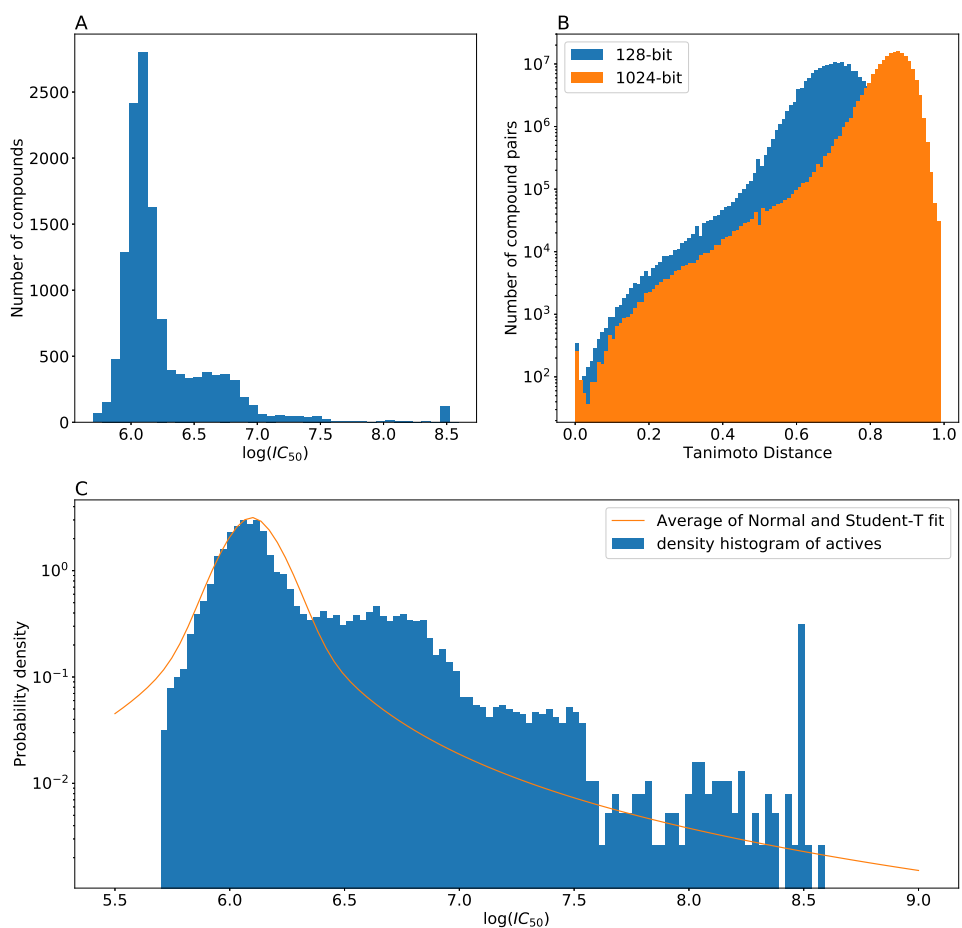}
\caption{{\bf Visual representation of the activity data in TCAMS.} Panel A: histogram of the distribution of the negative log (base 10) IC$_{50}$ of the compounds in the TCAMs data (n=13,533); Panel B: histogram of the distribution of pairwise Tanimoto distances between molecules in the TCAMs dataset under a 128-bit fingerprint representation (blue) and a 1024-bit fingerprint representation (orange); Panel C: the same density histogram as in panel A, but with the y-axis on a logarithmic scale, with the estimated mixture distribution used in the prediction procedure overlaid (average of a normal and a student-t distribution).}
\label{fig:cov}
\end{figure}

\subsubsection{Distance-dependent probability of being active}\label{stats_estimation}

The estimation of $P[x^* \in \mathcal{A} | d(x^*,\mathcal{L}_n)]$ is critical for the performance of the predictive model, see Equation \eqref{AnalysisGoal}.
This probability is proportional to the functional: \begin{equation}f_{n,N}(\delta) = \frac{P\left[d(x, \mathcal{L}_n)=\delta | x \in \mathcal{A}\right]}{P\left[d(x, \mathcal{L}_n)=\delta\right]}\end{equation} 
where $\delta \in [0, 1]$.  
An estimate $\hat{f}_{n,N}(\delta)$ of this functional should satisfy two properties:
\begin{enumerate}\label{sanity_checks}
    \item For $\delta=0$: $$ \hat{f}_{n,N}(0) = \frac{1-\epsilon}{P(x^* \in \mathcal{A})}$$ where $\epsilon << 1$ and depends on the granularity of the metric over molecular space.
    \item $\hat{f}_{n,N}(\delta)$ is monotonically decreasing in $\delta\in[0,1]$.
\end{enumerate}

To estimate $\hat{f}_{n,N}(\delta)$: (i) we generate random samples from the distribution $P\left[d(x, \mathcal{L}_n)=\delta | x \in \mathcal{A}\right]$ (the numerator); (ii) we generate random samples from the distribution $P \left[d(x, \mathcal{L}_n)=\delta\right]$ (the denominator); (iii) we use these two sets of random samples to determine a smooth estimate of the ratio as a function of $\delta$, such that the two properties specified above are satisfied.  
In this procedure, $\gamma$ is the bandwidth parameter of the Gaussian kernel density  used to estimate both probability densities for every value of $\delta$ (sklearn KernelDensity with default parameters). 

The optimal value of $\gamma$ is chosen as follows. 
First, we use the $v$-fold cross-prediction method to sample from $P\left[d(x, \mathcal{L}_n)=\delta | x \in \mathcal{A}\right]$ with $v=2$ and $k=5$, giving a total of 66635 samples (input to the numerator estimation).
Second, we choose ten equally spaced distances $\delta$ in the range $[0.. 0.45]$. 
For each of these distances $\delta$ we choose 10 samples of 100,000 points from the MolPort database using a specific sampling strategy explained below. 
We then use binary search to find the optimal bandwidth $\gamma$ such that the estimated $\hat{f}_n$ satisfies the property $\hat{f}_n(0) = 1.$.

This results in one hundred values for $\gamma$, and we take the median estimate $\hat{\gamma}$. We then use this $\hat{\gamma}$ to choose a value of $\delta$ such that samples chosen using this probability weighting, when smoothed with bandwidth $\gamma$, have $f_n(0) = 1$.  This gives us values (rounded) of $\gamma$ (bandwidth) = 0.09 and $\delta$ (for use in our sampling strategy) = 0.15.

The structure of the Molport data $\mathcal{U}_N$, whereby compounds early on in the numbering are much more likely to be close to the TCAMS dataset than those further on in the numbering motivates the following importance sampling type approach to choosing an appropriate subset of the data to use in fitting our estimate of $P \left[d(x, \mathcal{L}_n)=\delta\right]$. 
We generate sets of unlabelled data from $\mathcal{U}_N$, whereby the sampling probability decays as a function of the index of the unlabelled data using the following crude approach. The Molport data is divided into fifteen files, in increasing order (with 500,000 compounds per file, apart from the last which only has half this amount).  For a given distance value $\delta$, our sampling strategy goes as follows.  We calculate the number of compounds with minimum distance $\delta$ to the TCAMS data set, giving us $n_{\delta, i}$ for $i \in [0..14]$.  We sample from file $i$ (without replacement) with probability $n(\delta,i)/\sum_{j}(n_{\delta, j})$.

Throughout this paper we  used the python library scikit-learn \cite{scikit} version 0.19.1 and functions with default parameter settings except where stated otherwise.

\subsubsection{Degradation of predictive accuracy} 

To calculate the distance dependent degradation functions $\hat{\beta}(\delta), \hat{\epsilon}(\delta)$ (Equation \ref{penalty}), we choose a uniform grid of 10 values of $\delta$ spanning the interval [0,1].
For each $\delta$ value on this grid, we calculated $\hat{\beta}(\delta), \hat{\epsilon}(\delta)$ as per Equation \eqref{penalty} where the underlying regression models were random forests (RF) and ridge regression, respectively. 
We then used these ten estimates to interpolate smooth functions $\hat{\beta}(\delta)$ and $\hat{\epsilon}(\delta)$ by minimizing least squares deviation. The function is of the form $g(\delta) = a/(1 + e^{-b\delta^c})$. This function $g$ is continuous, strictly decreasing and non-negative over the interval $[0, 1]$, with three free parameters $(a,b,c)$.

\subsubsection{Testing of predictive models}\label{scores}

In order to benchmark the performance of the proposed predictive framework with respect to simpler alternatives, we designed testing experiments.  
Training and testing data were selected on the basis of quantiles of the distribution of the activity values \cite{watson:19}. In this set-up, all labelled data with activity values below a chosen activity quantile $q_{\text{train}}$ are used as training data, and all labelled data with activity values above a chosen activity quantile $q_{\text{test}}$ are used as part of the testing data. In particular, $q_{\text{train}} \leq q_{\text{test}}$. The complete testing set is then compose of these labelled data in addition to a set of unlabelled data. 

The thresholds used were $q_{\text{train}} = \{7.0,7.5\}$, and $q_{\text{test}}=\{7.5,8.0\}$. In the TCAMS dataset, there are 237 compounds with activity $\geq 7.5$, and 170 compounds with activity $\geq 8.0$. 
We denote $X_{q_{\text{train}}}$ as the training data defined by the cut-off $q_{\text{train}}$.
We denote $\hat{M}({\cdot | X_{q_{\text{train}}}})$ as the predictive model (random forest or ridge regression) fit to the training data $X_{q_{\text{train}}}$. 

Each compound $x^*$ in the testing data is ranked according to the following four scores:
\begin{enumerate}\label{test_methods}
    \item $S_0(x^*) = \hat{M}(x^* | X_{q_{\text{train}}})$. This is the predicted mean value of $y^*$. This is the unadjusted base model.
    \item $S_1(x^*) = \hat{\beta}\left[d(x^*,X_{q_{\text{train}}}\right] S_0(x^*)$. This is the predicted mean value of $y^*$ scaled by the distance-dependent penalty factor $\hat{\beta}(\delta)$, where $\delta$ is the setwise distance of $x^*$ from the training data. 
    \item $S_2(x^*) = P\left[x^*\in \mathcal{A} | d(x^*,X_{q_{\text{train}}})\right] S_1(x^*)$. This scores uses the additional reduction factor which is the probability that $x^*$ is active given its distance from the training data.
    \item $S_3(x^*) = F\left[S_2(x^*), \sigma^2\left(d(x^*,X_{q_{\text{train}}})\right)\right] P\left[x^*\in \mathcal{A} | d\left(x^*,X_{q_{\text{train}}}\right)\right]$, where $F(\mu,\sigma,\lambda)$ is the predicted cumulative distribution function of $y^*$ with mean $\mu$ and variance $\sigma$. This is the full model as specified in Equation \eqref{AnalysisGoal}.
\end{enumerate}

Figure \ref{fig:cov} shows the observed distribution of activities, which has a heavier tail than a Gaussian distribution. A Gaussian approximation of the observed activities gives a mean value of 6.25 and a standard deviation of 0.4, which implies that the expected number of compounds in the TCAMS dataset with activity $\geq 8$ is $0.08$, whereas in fact there 170.

For the cumulative distribution function $F$ in $S_3$, we choose a mixture model which is a combination of a normal and a student t-distribution (shown in panel C of Figure \ref{fig:cov}). We use the standard scikit-learn functions to fit a normal distribution to the activity data, and a Student-T distribution to that same data. Our mixture model is then simply the average of these two distributions. \footnote{This is an extremely crude way of fitting a Normal and Student-T mixture distribution, but as shown in Figure \ref{fig:cov} Panel C, it suffices to capture the fact that activity distribution has a long right tail, while also capturing the bulk of the distribution.}  We use this same distribution, but with the new values of $\mu$ and $\sigma$ to do our calculations for $S_3$. 
We implemented this fit using the inbuilt scipy fit functions, which fit distribution parameters to data.  We took as our model the average of the Normal fit to the activity data and the Student T-distribution fit to the data.  The approximation is shown against the density histogram in the bottom panel of Figure \ref{fig:cov}.

Finally, we choose our unlabelled data in one of two ways: `well-specified' and `mis-specified'. This corresponding to choosing a set of unlabelled compounds using the sampling method described above, which are closer or further to the TCAMS data, respectively. 
In each case we choose $500,000$ unlabelled compounds.  We use the same methodology as that used in calculating the fraction of actives to select the 'near' dataset (recall, this consists in choosing from each file according to the number of compounds with minimum distance 0.19 from the TCAMS dataset).  The `far' dataset is chosen in the same way, but the fraction chosen from each file is the inverse of the number of compounds at that distance.  This selection methodology aims to thus test the sensitivity of our results to the type of unlabelled data that the algorithm is searching over.

\subsubsection{Limitations of methodology}

A major limitation in the currently described methodology is that there is no propagation of uncertainty between the independent estimates. Further work would put this process into a fully Bayesian framework with uncertainty propagation.
In addition, the solution to the estimation of the ratio $f_{n,N}(\delta)$ is only approximate and could possibly be improved.

\subsection{Molecular Representation}

We standardized all chemical structures in all data sets described above to a common representation scheme using the python module standardizer (https://github.com/flatkinson/standardiser). Inorganic molecules were removed, and the largest fragment was kept in order to filter out counterions\cite{Fourches2010}. 
To represent molecules for subsequent model generation, we computed circular Morgan fingerprints\cite{Rogers2010} for all compounds using RDkit (release version 2013.03.02)\cite{rdkit}.
Specifically, we computed hashed Morgan fingerprints in binary format
using the RDkit function \textit{GetMorganFingerprintAsBitVect}, which returns values in $\mathbb{F}_2^{128}$,
and in count format, using in this case the RDkit function, \textit{GetHashedMorganFingerprint}, which returns values in $\mathbb{N}^{128}$.  

We decided to use Morgan fingerprints as compound descriptors given the higher retrieval rates obtained with this descriptor type in comparative virtual screening studies\cite{Koutsoukas2013}. The radius was set to 2 and the fingerprint length to 128. We note that longer fingerprints are associated with higher predictive power\cite{OBoyle2016}. However, a longer fingerprint of length 1024 did not provide a large improvement in terms of the activity covariance (Equation \ref{eq:active_sigma}) in these data as shown by Figure \ref{fig:ingredients}C. Hence, we decided to use the 128 fingerprint which is less likely to overfit.

\subsection{Data and Code Availability}

The code required to download directly from ChEMBL all these data sets, as well as the assay IDs for all of them, is available on the  accompanying GitHub repository for this article: https://github.com/owatson/PenalizedPrediction.

\section{Results}

\subsection{Semi-supervised framework for predicting highly active compounds}

Using Bayes rule, conditioning on (i) whether a novel compound is active and (ii) on the setwise distance between the compound and the training data, we formulate a predictive framework which adjusts for the selection bias inherent in many structure activity datasets. 
The rationale for this approach is that the labelled data - the available structure-activity data - are highly biased due to reporting selection bias (only reporting compounds with an activity level greater than some cutoff). In addition we take into account the empirical degradation of predictive performance as a function of the distance between the testing compound and the training compounds.
The goal is to obtain a model that can make unbiased predictions of activity for a previously unseen compound, explicitly adjusting for the degradation in predictive performance as a function of the distance to the training data.
This framework requires three elements. Firstly, a  metric over the space of small molecules whereby the distance between compounds explains a significant proportion of the covariance between their activities (in our case more than 50\%).
Secondly, we require a set of unlabelled (no corresponding activity measurements) compounds which are assumed to have been sampled under approximately the same data generating process as the labelled compounds, but without activity dependent reporting bias.
Thirdly, we require an estimate of the background frequency of the discovery of active compounds under the data generating process.

The framework does the following:
\begin{enumerate}
    \item We determine the probability of being active as a function of the distance to the training data, as given by Equation \eqref{eq:distance_dependent_activityprob}.
    \item We determine how the predictive accuracy of the model degrades as a function of the distance to the training data (Equation \eqref{penalty}).
    \item Determine how the covariance of the activity of two active elements varies as a function of the distance between them.
    \item Given some model for the full distribution of activity values of the active compounds (as a function of variance and expected activity level) - put the above three steps together to compute, for any unknown compound, the full posterior distribution of its activity.
\end{enumerate}

\begin{figure}
    \centering
    \includegraphics[width=\linewidth]{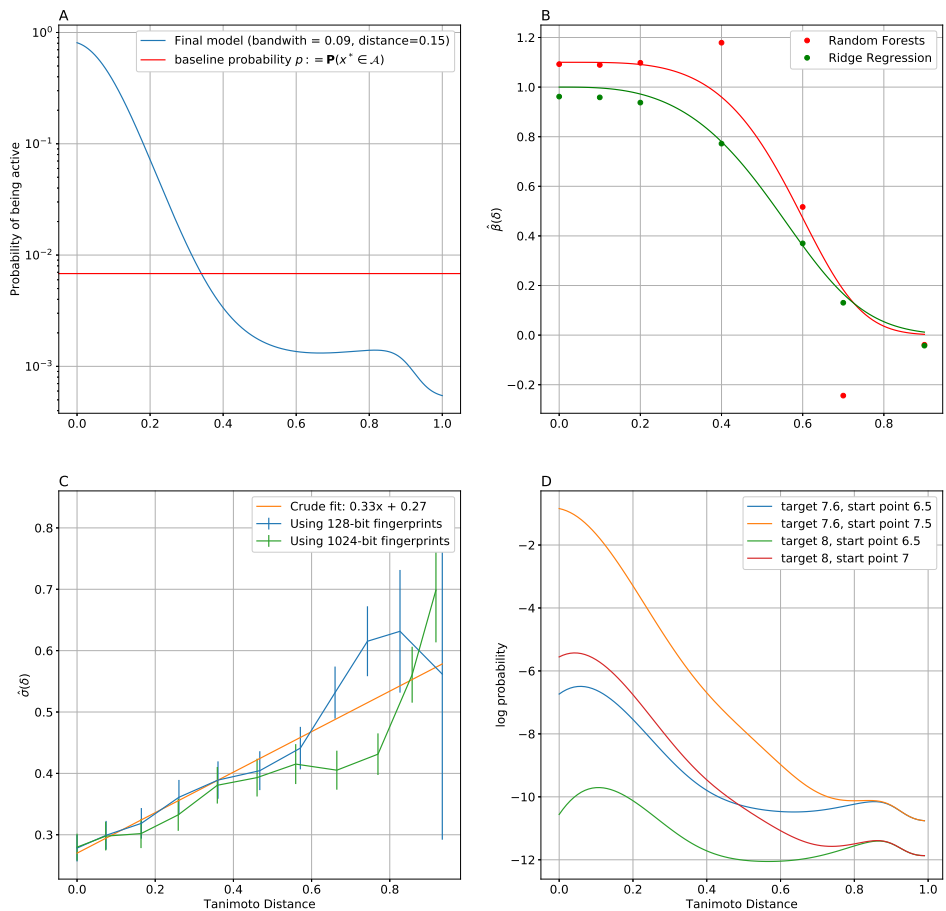}
    \caption{Overview of the model ingredients used for the adjusted predictions for the score $S_3$. A: the estimate of the fraction of compounds that are active as a function of the minimum distance to a known active. B: $\hat{\beta}(\delta)$ for Random Forests (equation \ref{penalty}) and Ridge Regression at various values of $d$, together with the smooth interpolation used. C: non-parametric estimation of the Tanimoto distance dependent activity covariance for both fingerprint representations (equation \ref{conditionalpredictivedist}).
    D: Plot of the contour lines of the log probability of finding a target compound of activity $Z$ at distance $d$ from a starting compound of activity $W$.  In all panels the x-axis is Tanimoto distance from the training set.}
    \label{fig:ingredients}
\end{figure}

In Figure \ref{fig:ingredients}, we illustrate the above four steps. 
Panel A shows the plot of the estimated probability that an unknown compound is active as a function of its distance $\delta$ to the training set comprised of only active compounds. Panel B plots the estimate of the degradation of predictive accuracy (model strength) as a function of $\delta$, for the two types of models we examine (ridge and random forest). Panel C shows how the covariance of two active compounds increases the further apart they are. This distance-dependent covariance is used to estimate the variance (and thus the standard deviation) of the activity of an unknown compound as a function of its distance to the known compounds in the training data. In addition, panel C shows how the observed covariances for the 128-bit fingerprint and the 1024-bit fingerprint compare: only marginal improvements are made with a more complex fingerprint representation.

Panel D merges these components and illustrates how the fully adjusted model (score $S_3$) works. We confine our attention to the random forest model. Rather than trying to plot the full predictive distribution as a function of $\delta$ for some compound (which would thus be a surface), we plot probability contour lines.  Given some unknown compound $x$, at distance $\delta$ to the active training set, suppose that $S_0(x) (= \hat{M}(x^* | X_{q_{\text{train}}}))$ is the simple estimate from the random forest model (without any adjustment of any kind) for the activity of $x$. We call this value the `start point'. Given some target level of activity $T$, we wish to plot the log probability that $y(x) >= T$ as a function of $\delta$.  We compute the probability that $x \in \mathcal{A}$ as a function of $\delta$. Then, assuming $x \in \mathcal{A}$, we compute the distribution of $y(x)$.  For this, we require three items:
\begin{enumerate}
    \item $\mu(\delta) := E\left[y(x)) | x \in \mathcal{A}\right]$. This is the score $S_1(x)$, which is  $S_0(x)$ adjusted towards the mean activity level as a function of $\delta$.
    \item $\sigma(\delta) := E\left[y(x) - y(a)\right]^2$ where $a \in \mathcal{A}$ is the closest compound to $x$.  We obtain this from the potency covariance plot (bottom left in Figure \ref{fig:ingredients}) as a function of $\delta$.
    \item The distribution of $y(x) | x \in \mathcal{A}$ as a function of $\mu(\delta)$ and $\sigma(\delta)$.  Here we use the distribution fit shown in the lower panel of Figure \ref{fig:cov}, but with our new estimates of $\mu(\delta)$ and $\sigma(\delta)$.  Once we have the distribution, we can read off the probability mass that lies above $T$.
\end{enumerate}

\subsection{Application to \textit{Plasmodium falciparum} screening data}

We analysed structure activity data on 13,533 compounds that were selected on the basis inhibiting \textit{P. falciparum} 3D7 growth by more than 80\% at 2 microMol \cite{Gamo2010}. 
To assess the benefit of using the semi-supervised framework, we compared the predictive performance between the derived semi-supervised predictive model (score $S_3$) and the standard fully supervised predictive model that does not use the unlabelled data (score $S_0$). Scores $S_1$ and $S_2$ are intermediate versions of the semi-supervised framework. The comparison between predictive frameorks (i.e. scores) was done using quantile-activity splitting \cite{watson:19}. This uses all compounds with activity below a certain threhsold as training data, and all compounds with activity above a certain threhsold as testing data.

We fit random forests and ridge regression models to two separate training sets: all compounds with activity less than 7 pIC50 and all compounds with activity less than 7.5 pIC50.
Two separate testing sets were used: all compounds with activity greater than 7.5 pIC50 (n=237), and all compounds with activity greater than 8 pIC50 (n=170). 
The predictive performance of each fitted model was then assessed under four different predictive frameworks (scores $S_0$ to $S_3$, see section \ref{scores}).

\begin{figure}
    \centering
    \includegraphics[width=\linewidth]{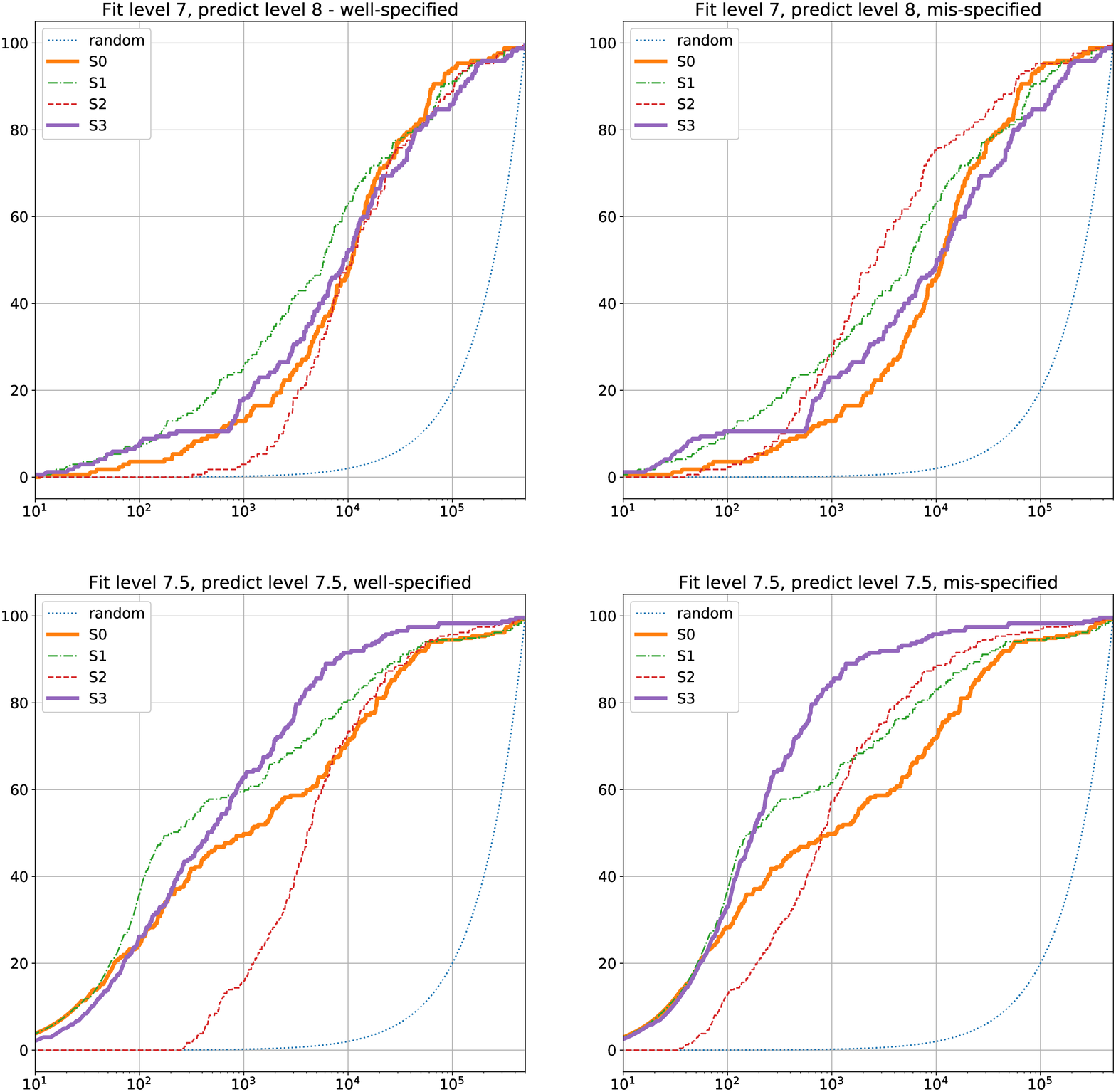}
    \caption{Comparison of predictive scores whereby random forests is the underlying predictive model.  Here the y-axis is $\%$ of active compounds found within the first $x$ compounds ordered by the selection methodology. }
    \label{fig:rf_perf}
\end{figure}

\begin{figure}
    \centering
    \includegraphics[width=\linewidth]{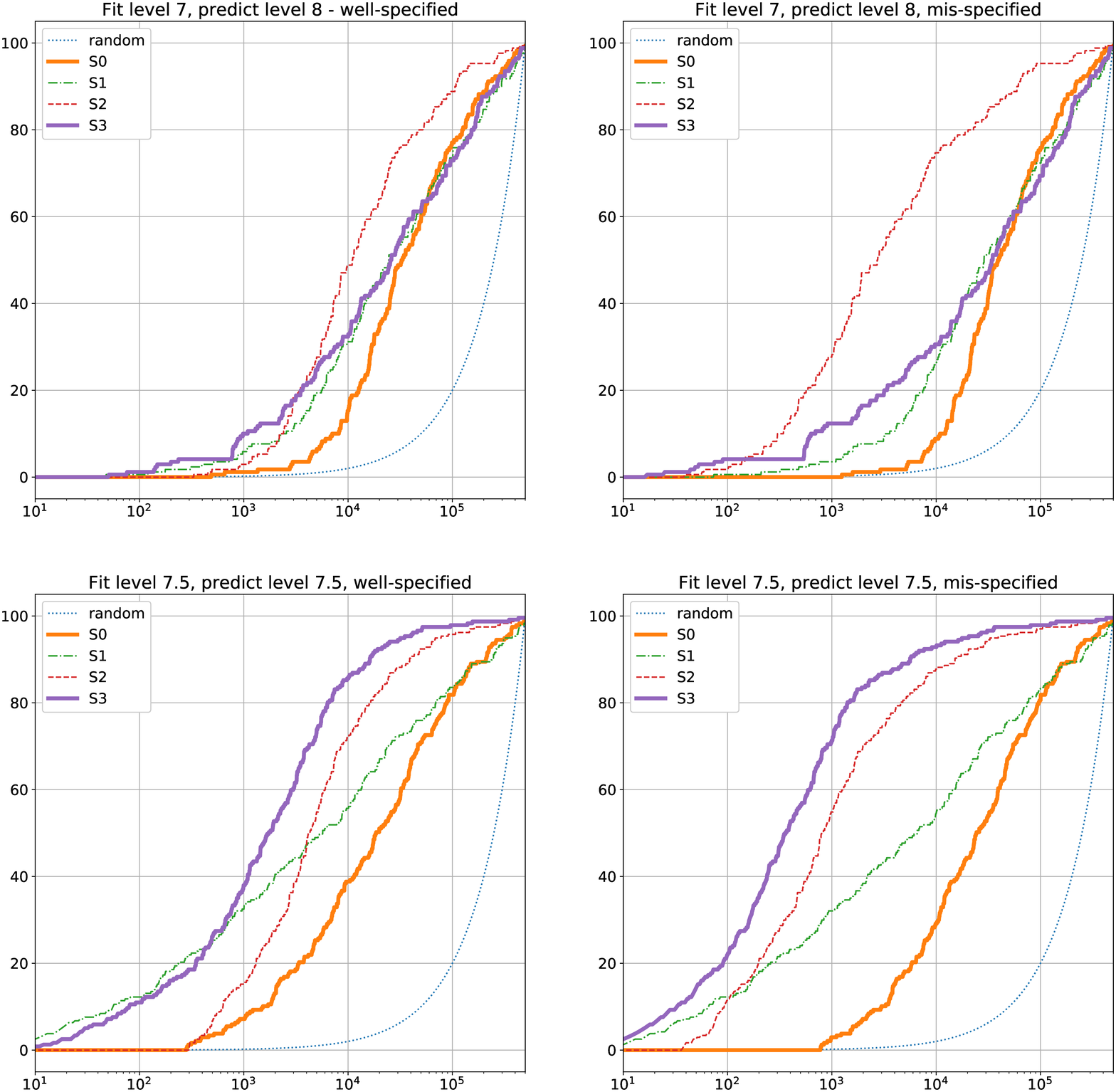}
    \caption{Comparison of predictive scores whereby ridge regression is the underlying predictive model Here the y-axis is $\%$ of active compounds found within the first $x$ compounds ordered by the selection methodology.}
    \label{fig:rdg_perf}
\end{figure}

A comparison of these four predictive frameworks is shown in Figure \ref{fig:rf_perf} for random forests and in Figure \ref{fig:rdg_perf} for ridge regression. For simplicity we show the results when training on compounds with activity less than 7 and testing on compounds greater than 8 (upper panels); and when training on compounds with activity less than 7.5 and testing on compounds with activity greater than 7.5 (lower panels). 
Each panel shows the percentage of true compounds (compounds in the TCAMS data not used in the model training stage and known to have activity above the desired threshold) discovered as a function of the number of compounds chosen from the testing set (500 000 compounds in total).
For a choice of 1000 compounds - a reasonable size for a drug discovery project - the naive model (score $S_0$) performs consistently worse across all experiments that the full predictive framework (score $S_3$).
For example, in the most difficult testing scenario, where the training data are all compounds with activity less than 7, and the testing compounds are those with activity greater than 8, then $S_3$ selects more active compounds than $S_0$ in the first $N$ compounds for $N$ up to around $10,000$ when the underlying model is random forests, and even more so when the underlying model is ridge regression.

\section{Discussion}

Our goal in this paper was to build QSAR models that fulfil two key goals. Firstly, for any given testing compound, the predicted activity is ``sensible''. By sensible we mean that the prediction takes into account the distance dependent degradation. This implies that for testing compounds whose structures are entirely different to all training compounds, the model's prediction will be based on the background discovery rate of active compounds and the mean activity of the active compounds.
Secondly, the model predictions should be ``useful''. By useful we mean that the adjusted model should outperform a `naive' model at distinguishing `good' compounds. We use a quantile-activity split approach to set up model testing experiments.
These two goals appear to be well aligned but they are not easy to jointly satisfy. For instance, the non-adjusted (`naive') random forest model (score $S_0$), only using the labelled data, performs almost as well as the fully adjusted model (score $S_3$) in identifying high-activity compounds in the training data (Figure \ref{fig:rf_perf}). 
However, the non-adjusted model does not make sensible predictions overall, since it predicts a non-negligible asexual activity against \textit{P. falciparum} 3D7 for any input compound.
Method $S_2$ does make sensible predictions by correctly predicting the average activity values for all compounds (due to the distance-dependent adjustment), but underperforms $S_0$ substantially in three of our four testing experiments setups.
Indeed, a simple example of a sensible, but completely useless model is one that predicts the background adjusted mean activity level for any compound.

We show that these two goals can be achieved by explicitly modelling the full distribution of our prediction, rather than just the mean value, and taking this distribution into account in the optimization process.  The method that does this ($S_3$ in Figures \ref{fig:rf_perf} and \ref{fig:rdg_perf}) is the top performing method for choosing compounds overall. It is the top performing method in four of the eight tests performed, and no other method consistently dominates it (the closest is method $S_1$, which, like $S_0$, does not make sensible predictions overall).

The utility of having a general predictive model framework that performs both these goals is that it opens up new questions for quantitative analysis, and in particular optimization.  
For optimization algorithms to converge, they need not only to produce accurate answers on the domain of interest (what we call a `useful' model), but they also need to provide at least approximately correct answers outside that domain (what we call a `sensible' model).  
In our testing experiments, all the methods tested ($S_0$ to $S_3$) provide rankings of all compounds. 
However the fully adjusted model (score $S_3$) has an additional advantage. The rank it provides for a given compound is derived from the probability that the compound will an activity above a threshold of interest.  Thus given three compounds $x_0, x_1, x_2$, with $S_3(x_0) > S_3(x_1) > S_3(x2)$, we can ask the question `would we have a higher chance of finding at least one compound with an activity above the threshold of interest if we tested $x_1$ and $x_2$, rather than just $x_0$?  This question cannot be answered by the other model adjustments, and this example can of course be extensively generalized.  Most of the practical questions that face researchers in this area are in terms of tradeoffs, e.g. ``how many compounds should we make in one batch?''; ``how similar should they be?''; ``is it work making one expensive compound that is predicted to be highly active, or testing ten cheap ones that are not predicted to be quite as good?''\cite{Valler2000,Huggins2011}. We hope that this approach will make predictive models substantially more useful to practitioners.

\section{Author Contributions}
O.W. designed the study and analysed the data. 
I.C.C curated the dataset.
J.W. wrote the mathematical framework for the model. 
O.W. and J.W. interpreted the results, and wrote the paper.
All authors read and approved the final manuscript.

\section{Acknowledgements}
We would like to thank Molport for making their proprietary screening library available to us for our research.

\section{Conflicts of interest}
O.W. and I.C.C. hold equity interest in Evariste Technologies Ltd.

\bibliographystyle{unsrtnat}
\bibliography{references}
\end{document}